\documentclass{moriond}

\usepackage{subcaption}
\usepackage{graphicx}
\usepackage{siunitx}
\usepackage{physics}

\def\be{\begin{equation}}
\def\ee{\end{equation}}
\def\bea{\begin{eqnarray}}
\def\eea{\end{eqnarray}}

\newcommand{\squark}{{\tilde q}}
\newcommand{\upsquark}{{\tilde u}}

\newcommand{\gluino}{{\tilde g}}
\newcommand{\gaugino}{{\tilde \chi}}

\newcommand{\Photo}{\includegraphics[height=35mm]{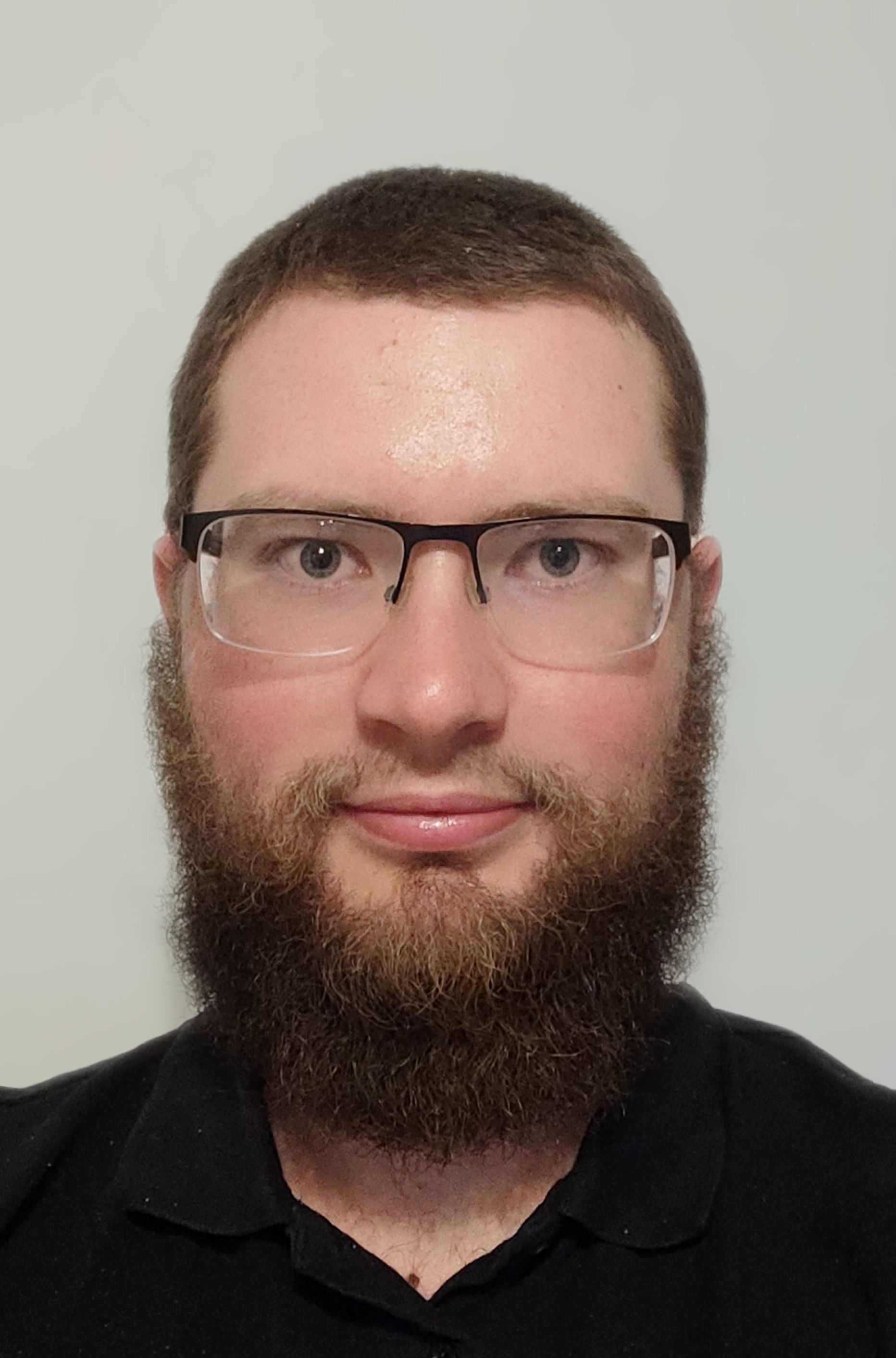}}

\begin{document}

\begin{flushright}
       MS-TP-22-14\\
\end{flushright}

\vspace*{4cm}
\title{Associated squark-electroweakino production with NLO+NLL precision}

\author{Juri Fiaschi$^{1}$, Benjamin Fuks$^{2}$, Michael Klasen$^{3}$, Alexander Neuwirth$^{3}$ \footnote{Speaker} }

\address{
       $^1$Department of Mathematical Sciences, University of Liverpool, Liverpool L69 3BX, United Kingdom \\
       $^2$Sorbonne Université,  CNRS,  Laboratoire de Physique Théorique et Hautes Énergies, LPTHE, F-75005 Paris, France \\
       $^3$Institut  für  Theoretische  Physik,  Westfälische  Wilhelms-Universität  Münster,\newline  Wilhelm-Klemm-Straße 9, 48149 Münster, Germany}

\maketitle\abstracts{
       Motivated by the increased precision expected from LHC Run 3, equally accurate theory predictions are mandatory.
       As supersymmetry mass limits increase, predictions can be improved by threshold resummation.
       We examine the effects of including next-to-leading logarithms on associated squark-electroweakino production at the LHC and find a significant reduction in the uncertainty of factorisation and renormalisation scale dependence and a modest increase in the total cross section.
}
\section{Introduction}

Supersymmetry (SUSY) is a prominent extension of the Standard Model (SM) and will be further investigated in the upcoming LHC Run 3.
The Minimal Supersymmetric Standard Model (MSSM) addresses some open questions in the SM, as it protects the mass of the Higgs boson from radiative corrections, predicts the unification of strong and electroweak forces at high scales, and also includes Dark Matter candidates.
Since light SUSY particles are excluded by direct searches at the LHC and if squarks and gluinos are found to be too heavy to be produced in pairs, the associated production of a squark or gluino with an electroweakino becomes important.
The semi-strong production of one strongly and one electroweak-interacting superpartner offers cross sections of intermediate size and a larger available phase space due to having a typically lighter electroweakino in the final state.
However, current mass limits also imply that in any SUSY production process, the kinematic configuration approaches the production threshold. 
This results in large threshold logarithms ruining the convergence of the perturbative series, so that they must be resummed.
Soft-gluon resummation accounts for these logarithms to reduce otherwise fairly large theoretical uncertainties.
Precise predictions of strong \cite{Kulesza:2008jb,Beenakker:2009ha,Beenakker:2011sf,Beenakker:2014sma,Beenakker:2016gmf} and electroweak \cite{Bozzi:2007qr,Debove:2010kf,Fuks:2013vua,Fuks:2016vdc,Fiaschi:2019zgh} SUSY processes beyond NLO have been achieved in the last decade.
We briefly review the threshold resummation formalism that can be used for such precision calculations in section \ref{subsec:resum} and illustrate their main effects in section \ref{subsec:proc} for squark-electroweakino production.
We summarize our work in Section \ref{subsec:summary}.

\section{Soft gluon resummation}\label{subsec:resum}
After the cancellation of soft and collinear divergences between real and virtual corrections, large logarithms will remain near the threshold region due to the different phase spaces \cite{Kinoshita:1962ur,Lee:1964is}.
Since these logarithms spoil the convergence of fixed order calculations, the emission of soft gluons up to all orders must be included.
To do this, the computation must factorize both dynamically, by using eikonal Feynman rules, and kinematically, by transforming into Mellin space.
Then, the large logarithms depend on the Mellin variable $N$ and the hadronic differential cross section $\dd \sigma_{AB}/\dd M^2$ in the conjugate $N$-space depends on the invariant mass $M$, parton densities $f_{i/h}$ and partonic cross section $\sigma_{ab}$ \cite{Collins:1989gx},
\begin{equation}
	M^2 \frac{\dd \sigma_{AB}}{\dd M^2} (N-1) = \sum_{a,b}\ f_{a/A}(N,\mu_F^2)\ f_{b/B}(N,\mu_F^2)\  \sigma_{ab}(N,M^2,\mu_F^2,\mu_R^2)
       \,.
\label{eq:HadFacN}
\end{equation}
The soft and collinear gluon radiation is embedded in the Sudakov form factors $G$.
In particular, for squark-electroweakino production $G^{(2)}_{ab \to ij}$ includes the process-dependent modified soft anomalous dimension which is closely related to the topologically similar production of $tW$ \cite{Kidonakis:2006bu}.
Then, the partonic cross section, depending on both the factorisation scale $\mu_F$ and the renormalisation scale $\mu_R$, is expressed in an exponential form scaled by the hard function $\mathcal H$,
\begin{equation}
	\sigma^{\text{Res.}}_{ab\to ij}(N,M^2,\mu_{F,R}^2) = \mathcal H_{ab \to ij}(M^2,\mu_{F,R}^2)\ \exp\Big[ \underbrace{LG^{(1)}_{ab}(N)}_{\text{LL}} + \underbrace{G^{(2)}_{ab \to ij}(N,M^2,\mu_{F,R}^2)}_{\text{NLL}} + \dots \Big]
	\label{eq:HtimesG}
\end{equation} 
truncated at next-to-leading logarithmic order (NLL).
Since squark-electroweakino production involves only one colour basis tensor, we dropped the irreducible colour representation index \cite{Beenakker:2013mva}.
To consistently include these logarithms, we subtract the resummed cross section $\sigma^\text{Res.}$ expanded to $\mathcal O (\alpha_s^2)$,
\begin{equation}
	\sigma_{ab} = \sigma_{ab}^\text{NLO} + \sigma_{ab}^{\text{Res.}}  - \sigma_{ab}^\text{Exp.}
       \,.
\end{equation}
This avoids double counting of $\mathcal O (\alpha_s^2)$ contributions that are already fully contained by the complete next-to-leading order calculation $\sigma^\text{NLO}$.
Due to singularities in the $N$-space cross section returning from Mellin space by an inverse transformation, a distorted integration contour following the principal value procedure and minimal prescription is required \cite{Contopanagos:1993yq,Catani:1996yz}.

\section{Squark-electroweakino production at next-to-leading logarithmic accuracy}\label{subsec:proc}
\begin{figure}
       \centering
       \begin{subfigure}[b]{\textwidth}
           \centering
\includegraphics[width=\linewidth]{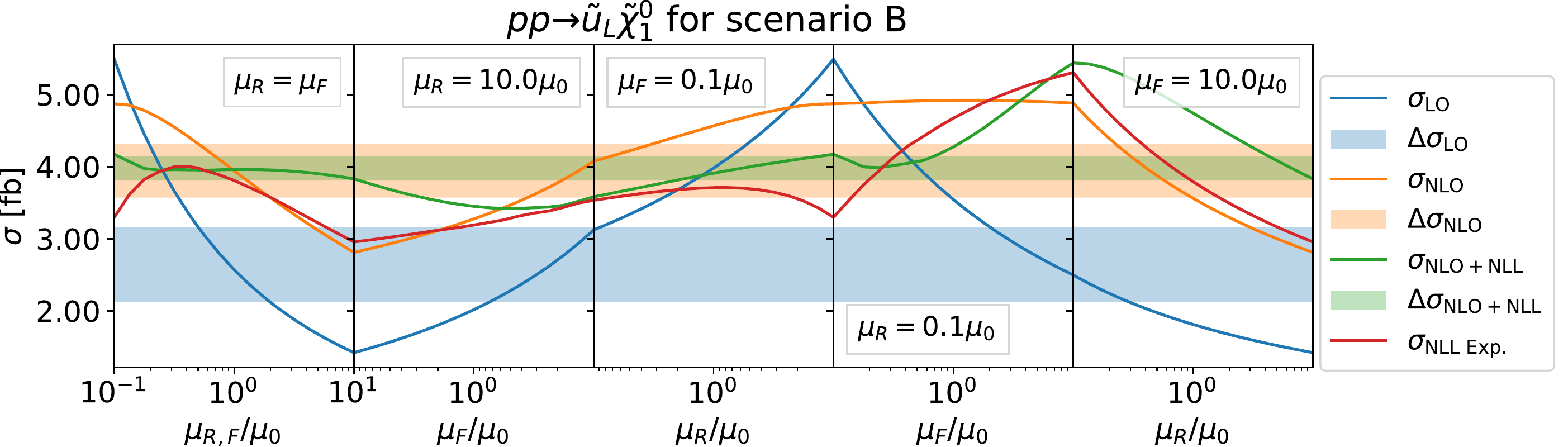}
       \end{subfigure}
	\caption{
	Profile of the renormalisation and factorisation scale dependence of the total cross section in the process $pp \to\tilde u_{L} \gaugino_1^0$ at $\sqrt{S}=\SI{13}{TeV}$. 
       The plots cover $\mu_{F,R} \in (0.1-10)\mu_0$ (reversed in panels 2 and 3) with the central scale $\mu_0 = (m_\squark +m_\gaugino)/2$.
    	The bands show the scale uncertainties from the seven-point method.
	}
       \label{fig:scale}
\end{figure}

\begin{figure}
       \centering
       \begin{subfigure}[b]{0.49\textwidth}
           \centering
            \includegraphics[width=\textwidth]{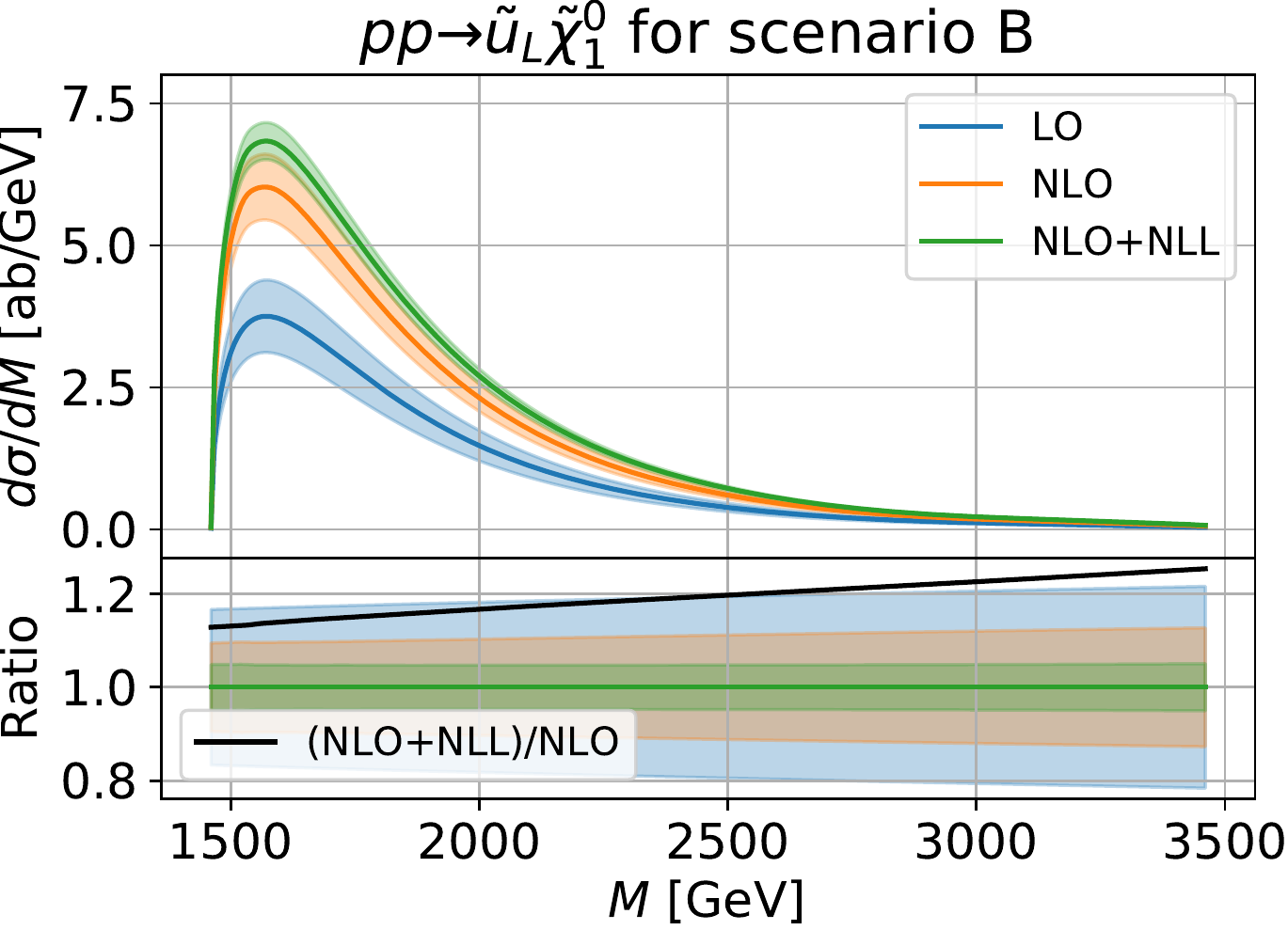}
            \caption{Invariant mass distribution.}
            \label{fit:inv_mass}
       \end{subfigure}
       \hfill
       \begin{subfigure}[b]{0.49\textwidth}
           \centering
            \includegraphics[width=\textwidth]{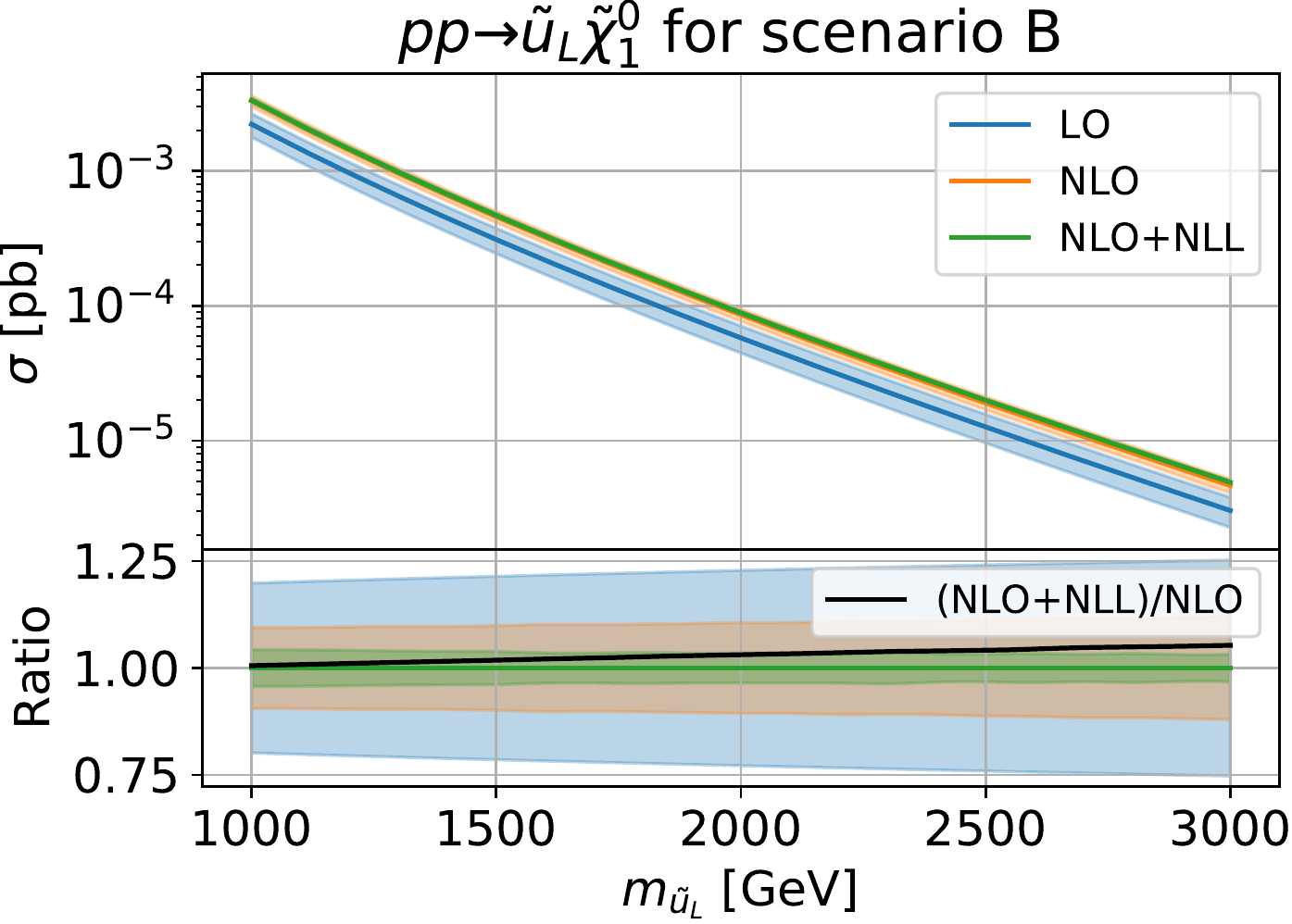}
            \caption{Total cross section by the squark mass $m_{\upsquark_{L}}$.}
            \label{fit:sq_mass}
       \end{subfigure}
	\caption{
		For the process $pp \to \upsquark_{L}\gaugino^0_1$ at $\sqrt S = \SI{13}{TeV}$ we show the mass dependence of the cross section.
              The lower panels show relative scale uncertainties and (NLO+NLL)/NLO $K$-factors.
	}
	\label{fig:mass}
\end{figure}

In Fig.~\ref{fig:scale} we focus on the scale variation of the total cross section for a pMSSM-11 scenario with $m_\squark=\SI{1}{TeV}$, $m_{\gaugino_1^0}=\SI{0.5}{TeV}$ and $m_\gluino = \SI{3}{TeV}$ \cite{Fiaschi:2022odp}.
The uncertainty bands are determined by the seven-point method, varying the factorisation and renormalisation scales independently by factors of 2 around the central scale $\mu_0 = (m_\squark + m_\gaugino)/2$, with relative factors of 4 excluded, which means $1/2 \leq \mu_R/\mu_F \leq 2$. 
For large scales, the logarithms become dominant and the expansion $\sigma^{\text{Exp.}}$ approaches the NLO $\sigma^{\text{NLO}}$.
We observe that the resummation does not significantly increase the cross section at the central scale, but it reduces the scale uncertainty.
For different scenarios with heavier superpartners the logarithms have an effect on the total cross section at central scale (cf. Fig.~\ref{fig:mass}).
Across several scenarios we observed that the relative scale uncertainty improved from about \SI{20}{\%} at LO to \SI{10}{\%} at NLO and finally at NLO+NLL below \SI{5}{\%}.

In Fig.~\ref{fig:mass} we first show the invariant mass distribution on the left.
As the invariant mass M increases, we approach the threshold region $M^2/s\to 1$, and the NLL corrections contribute significantly more to the differential cross section. 
This behaviour is captured by the NLL+NLO/NLO $K$-factors shown in the lower panels of the figure.
The lower panels also display the relative seven-point scale uncertainty, where the scale is varied around a central scale choice of $\mu_0 = M$.
The second figure shows the total cross section at $\mu_0 = (m_\squark + m_\gaugino)/2$.
While the central cross section values are enlarged by \SI{50}{\%} from LO to NLO, the additional increase from NLL resummation reaches only about \SI{6}{\%} in the mass ranges observable at the LHC in the near future.

\begin{figure}
       \centering
       \begin{subfigure}[b]{\textwidth}
           \centering
\includegraphics[width=0.5\linewidth]{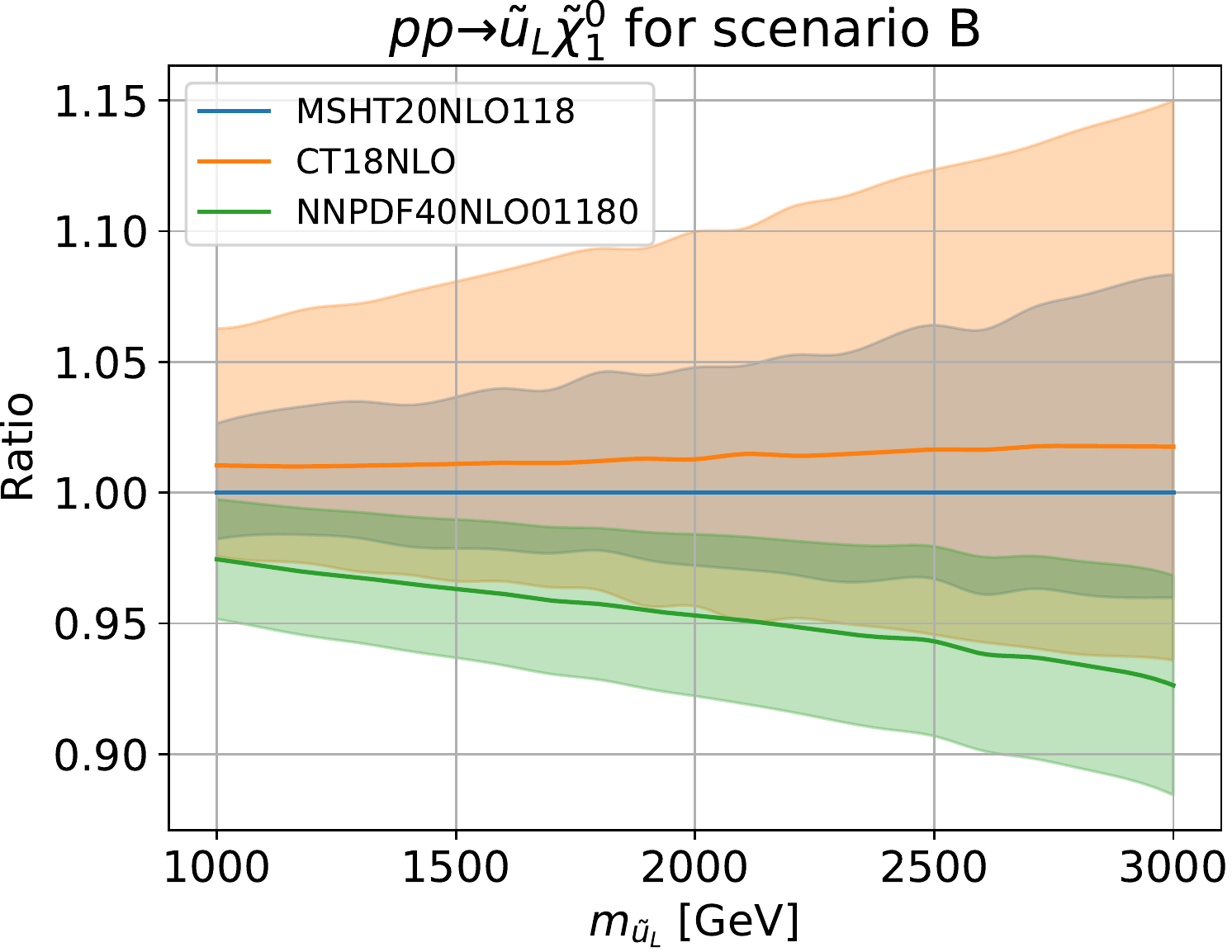}
       \end{subfigure}
	\caption{
              Relative PDF uncertainties at \SI{90}{\%} confidence level for the total cross sections of the process $pp \to\tilde u_{L} \gaugino_1^0$ at $\sqrt{S}=\SI{13}{TeV}$ and NLO+NLL. 
	}
       \label{fig:pdf}
\end{figure}

Fig.~\ref{fig:pdf} displays the uncertainties from the MSHT20, CT18 and NNPDF40 parton distribution functions.
While the uncertainty is of about \SI{5}{\%} for \SI{1}{TeV} squarks, it increases up to \SIrange{10}{15}{\%} for squark masses of \SI{3}{TeV}. 
The central cross section values obtained with the MSHT20 and CT18 sets agree consistently.
The NNPDF40 predictions are a few percent lower, but still in reasonable agreement within their uncertainty intervals.

\section{Summary}\label{subsec:summary}
We have presented a threshold resummation calculation with NLO+NLL accuracy for the associated production of a squark and an electroweakino at the LHC.
By matching fixed order and resummed predictions, we have consistently combined the resummation of large logarithms, which appear close to threshold, with NLO results.
The NLL resummation increased the NLO cross sections for central scales by up to \SI{6}{\%} and reduced the scale uncertainty to $\pm\SI{5}{\%}$.
Our calculation adds squark-electroweakino production to existing slepton pair, electroweakino pair and electroweakino-gluino  production in the public code \href{https://resummino.hepforge.org}{\textsc{Resummino} (resummino.hepforge.org)}.

\section*{Acknowledgments}

This work has been supported by the BMBF under contract 05P21PMCAA and by the DFG through the Research Training Network 2149 “Strong and Weak Interactions - from Hadrons to Dark Matter”.
A.N. thanks the organizers for the great conference.


\section*{References}

\begin{thebibliography}{99}

\bibitem{Kulesza:2008jb}
A.~Kulesza and L.~Motyka,
Phys. Rev. Lett. \textbf{102} (2009), 111802

\bibitem{Beenakker:2009ha}
W.~Beenakker, S.~Brensing, M.~Kramer, A.~Kulesza, E.~Laenen and I.~Niessen,
JHEP \textbf{12} (2009), 041

\bibitem{Beenakker:2011sf}
W.~Beenakker, S.~Brensing, M.~Kramer, A.~Kulesza, E.~Laenen and I.~Niessen,
JHEP \textbf{01} (2012), 076

\bibitem{Beenakker:2014sma}
W.~Beenakker, C.~Borschensky, M.~Kr\"amer, A.~Kulesza, E.~Laenen, V.~Theeuwes and S.~Thewes,
JHEP \textbf{12} (2014), 023

\bibitem{Beenakker:2016gmf}
W.~Beenakker, C.~Borschensky, R.~Heger, M.~Kr\"amer, A.~Kulesza and E.~Laenen,
JHEP \textbf{05} (2016), 153

\bibitem{Bozzi:2007qr}
G.~Bozzi, B.~Fuks and M.~Klasen,
Nucl. Phys. B \textbf{777} (2007), 157-181


\bibitem{Debove:2010kf}
J.~Debove, B.~Fuks and M.~Klasen,
Nucl. Phys. B \textbf{842} (2011), 51-85

\bibitem{Fuks:2013vua}
B.~Fuks, M.~Klasen, D.~R.~Lamprea and M.~Rothering,
Eur. Phys. J. C \textbf{73} (2013), 2480

\bibitem{Fuks:2016vdc}
B.~Fuks, M.~Klasen and M.~Rothering,
JHEP \textbf{07} (2016), 053

\bibitem{Fiaschi:2019zgh}
J.~Fiaschi, M.~Klasen and M.~Sunder,
JHEP \textbf{04} (2020), 049

\bibitem{Kinoshita:1962ur}
T.~Kinoshita,
J. Math. Phys. \textbf{3} (1962), 650-677

\bibitem{Lee:1964is}
T.~D.~Lee and M.~Nauenberg,
Phys. Rev. \textbf{133} (1964), B1549-B1562

\bibitem{Collins:1989gx}
J.~C.~Collins, D.~E.~Soper and G.~F.~Sterman,
Adv. Ser. Direct. High Energy Phys. \textbf{5} (1989), 1-91

\bibitem{Beenakker:2013mva}
W.~Beenakker, T.~Janssen, S.~Lepoeter, M.~Kr\"amer, A.~Kulesza, E.~Laenen, I.~Niessen, S.~Thewes and T.~Van Daal,
JHEP \textbf{10} (2013), 120
\bibitem{Kidonakis:2006bu}
N.~Kidonakis,
Phys. Rev. D \textbf{74} (2006), 114012

\bibitem{Contopanagos:1993yq}
H.~Contopanagos and G.~F.~Sterman,
Nucl. Phys. B \textbf{419} (1994), 77-104

\bibitem{Catani:1996yz}
S.~Catani, M.~L.~Mangano, P.~Nason and L.~Trentadue,
Nucl. Phys. B \textbf{478} (1996), 273-310

\bibitem{Fiaschi:2022odp}
J.~Fiaschi, B.~Fuks, M.~Klasen and A.~Neuwirth,
[arXiv:2202.13416 [hep-ph]].

\end{thebibliography}
\end{document}